\documentclass[a4paper,11pt]{article}
\usepackage{pos}

\usepackage{tikz,graphicx,booktabs,multirow,array,microtype,mathtools,float}

\usepackage{centernot}
\newcommand{\fsl}[1]{{\centernot{#1}}} 

\graphicspath{{figures/}}
\usepackage{enumerate}   
\extrafloats{200}
\usepackage{anyfontsize}
\usepackage[shortlabels]{enumitem}

\usepackage{amsfonts}
\AtBeginDocument{
\heavyrulewidth=.08em
\lightrulewidth=.05em
\cmidrulewidth=.03em
\belowrulesep=.65ex
\belowbottomsep=0pt
\aboverulesep=.4ex
\abovetopsep=0pt
\cmidrulesep=\doublerulesep
\cmidrulekern=.5em
\defaultaddspace=.5em
}

\newcommand{\langl}{\begin{picture}(4.5,7)
\put(1.1,2.5){\rotatebox{60}{\line(1,0){5.5}}}
\put(1.1,2.5){\rotatebox{300}{\line(1,0){5.5}}}
\end{picture}}
\newcommand{\rangl}{\begin{picture}(4.5,7)
\put(.9,2.5){\rotatebox{120}{\line(1,0){5.5}}}
\put(.9,2.5){\rotatebox{240}{\line(1,0){5.5}}}
\end{picture}}

\usepackage{amsmath} 
\usepackage{dsfont}  
\usepackage{bm}      

\def\beq{\begin{equation}}
\def\eeq{\end{equation}}
\def\beqs#1\eeqs{\beq\begin{split} #1 \end{split}\eeq}

\long\def\comment#1{}

\title{Measuring Charged Particle Polarizabilities on the Lattice without Background Fields}
\ShortTitle{Measuring Charged Particle Polarizabilities...}

\author[a]{Walter Wilcox}
\author[b]{Frank X. Lee}

\affiliation[a]{Department of Physics, Baylor University,\\
 Waco, Texas 76798, USA}

\affiliation[b]{Physics Department, The George Washington University,\\
 Washington, DC 20052, USA}

\emailAdd{Walter\_Wilcox@baylor.edu}
\emailAdd{fxlee@gwu.edu}

\abstract{We show how to compute electromagnetic polarizabilities of charged hadrons without the use of background fields in lattice QCD. 
The low-energy behavior of the Compton scattering amplitude is matched to matrix elements of current-current correlation functions on the lattice. Working in momentum space, formulas for electric polarizability ($\alpha_E$) and magnetic polarizability ($\beta_M$) are derived for both charged pion and proton. 
Lattice four-point correlation functions are constructed from quark and gluon fields to be used in Monte-Carlo simulations. We also draw attention to the potential of four-point functions as a multi-purpose tool for hadron structure.}

\FullConference{%
 The 38th International Symposium on Lattice Field Theory, LATTICE2021
  26th-30th July, 2021
  Zoom/Gather@Massachusetts Institute of Technology
}

\begin{document}
\maketitle

\section{\label{sec:intro}Introduction}

Understanding electromagnetic polarizabilities has been a long-term goal of lattice QCD. The challenge lies in the need to apply both QCD and QED principles. The standard tool to compute polarizabilities is the background field method which has been widely used~\cite{Fiebig:1988en,Lujan:2016ffj, Lujan:2014kia, Freeman:2014kka, Alexandru:2008sj, Lee:2005dq, Lee:2005ds,Engelhardt:2007ub}. 
Although such calculations are relatively straightforward, requiring only two-point functions, there are a number of unique challenges. These include the problem of removing electro-quenching of the external field and the fact that charged particles accelerate in an electric field and exhibit Landau levels in a magnetic field. For this reason, most calculations have focused on neutral hadrons.

In this work, we examine the use of four-point functions to extract polarizabilities (A fuller version is published in Ref.~\cite{Wilcox:2021rtt}). 
As we shall see, the method is ideally suited to charged hadrons. Although four-point correlation functions have been applied to various aspects of hadron structure~\cite{Liang:2019frk},  
not too much attention has been paid to its potential application for polarizabilities.
The only work we are aware of are two attempts 25 years ago, one based in position space~\cite{BURKARDT1995441}, one in momentum space~\cite{Wilcox:1996vx}. 

\section{Charged pion}
\label{sec:pion} 

\begin{figure}[h]
\begin{center}
\includegraphics[scale=0.50]{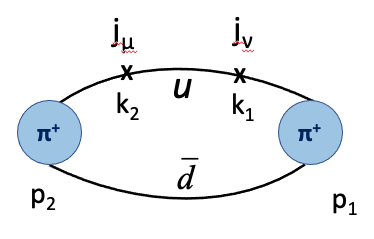}
\caption{Pictorial representation of the four-point function in Eq.\eqref{eq:4pt} for $\pi^+$ (for proton imagine two $u$ and one $d$ quark lines). 
Time flows from right to left and the four-momentum conservation is $p_2 +k_2 = k_1+p_1$.}
\label{fig:diagram-4pt1}
\end{center}
\end{figure}

We follow closely the notations and conventions of Ref.~\cite{Wilcox:1996vx}.
The central object is the time-ordered Compton scattering tensor 
defined by the four-point correlation function\footnote{We use round brackets $(\cdots |\cdots)$ to denote continuum matrix elements, and angle brackets $\langl \cdots |\cdots\rangl$ lattice matrix elements.}, 
\begin{equation}
T_{\mu\nu}=i\int d^4x e^{ik_2\cdot x}( \pi(p_2)|T j_\mu(x) j_\nu(0) | \pi(p_1))
\label{eq:4pt}
\end{equation}
where the electromagnetic current density $j_\mu=q_u \bar{u}\gamma_\mu u + q_d \bar{d}\gamma_\mu d$, is built from up and down quark fields ($q_u=2/3$, $q_d=-1/3$).
The function is represented in Fig.~\ref{fig:diagram-4pt1}.
We work with a special kinematical setup called the zero-momentum Breit frame given by,
\begin{eqnarray}
&p_1 =(m,\vec{0}),\nonumber\\
&k_1 =(0,\vec{k}),\;
k_2 =(0, \vec{k}), \; \vec{k}=k \hat{z},\;  k \ll m,\label{eq:Breit1}\\
&p_2 =-k_2+k_1+p_1=(m,\vec{0}),\nonumber
\end{eqnarray}
Essentially it can be regarded as forward double virtual Compton scattering.

On the phenomenological level, the process can be described by an effective relativistic theory to expose its physical content. 
The tensor can be parameterized to second order in photon momentum by the general form,
\begin{eqnarray}
&\sqrt{2E_12E_2} \,T_{\mu\nu} = 
 -{T_\mu(p_1+k_1,p_1) T_\nu(p_2,p_2+k_2) \over (p_1+k_1)^2-m^2 } 
 -{T_\mu(p_2,p_2-k_1) T_\nu(p_1-k_2,p_1) \over (p_1-k_2)^2-m^2 } \nonumber \\
&+2g_{\mu\nu}+ A(k_1^2g_{\mu\nu} - k_{1\mu}k_{1\nu} + k_2^2 g_{\mu\nu} - k_{2\mu}k_{2\nu}) 
+ B(k_1\cdot k_2 g_{\mu\nu} -  k_{2\mu}k_{1\nu}) \label{eq:Tmn}\\
&+ C(k_1\cdot k_2 Q_\mu Q_\nu + Q\cdot k_1 Q\cdot k_2  g_{\mu\nu} 
- Q\cdot k_2 Q_\mu k_{1\nu} - Q\cdot k_1 Q_\nu k_{2\mu}),\nonumber
\end{eqnarray} 
where $Q=p_1+p_2$ and $A$, $B$, $C$ are constants to be characterized. We use a non-covariant normalization 
\begin{equation}
\sum_n\int {d^3p\over (2\pi)^3} |n(p))(n(p)|=1,
\end{equation}
which is why the square root factor is in front of $T_{\mu\nu} $.
The pion electromagnetic vertex with momentum transfer $q=p'-p$ is written as 
\begin{equation}
T_\mu(p',p)=(p'_\mu+p_\mu) F_\pi(q^2) +q_\mu {p'^2-p^2\over q^2}(1-F_\pi(q^2)).
\end{equation}
It satisfies $q_\mu T_\mu(p',p)=p'^2-p^2$ for off-shell pions, which is needed for the Ward-Takahashi identity. Current conservation ($k^\mu_1T_{\mu\nu}= k^\nu_2T_{\mu\nu}=0$) immediately leads to $A$ being related to the charge radius by $A= {\langl r^2\rangl / 3}$.
The first three terms on the right in Eq.\eqref{eq:Tmn} are the Born contributions to scattering from the pion and the remaining three are contact terms. The electric polarizability, $\alpha_E$, and magnetic polarizability, $\beta_M$, terms come from $B$ and $C$,
\begin{equation}
\alpha_E \equiv -\alpha\left( {B\over 2m} + 2mC \right),
\beta_M \equiv  \alpha {B\over 2m}.
\label{eq:BC}
\end{equation}
For electric polarizability, we work with the $\mu=\nu=0$ component of Eq.\eqref{eq:Tmn}. To order $ \vec{k}^{\,2}$ one has
\begin{equation}
T_{00}=T^{Born}_{00}+  {\alpha_E \over \alpha}  \vec{k}^{\,2}.
\end{equation}

The next step is to relate the polarizabilities to lattice matrix elements.
To this end, we need to convert from continuum to a lattice of isotropic spacing $a$ with $N_s=N_x\times N_y\times N_z$  number of spatial sites  by the following correspondence,
\begin{eqnarray}
|n(p)) \to V^{1/2} |n(p)>, \,\,
j_\mu(x) \to {Z_V \over a^3} j^{L}_\mu(x), \,\,
\int d^4x \to a^4 \int_{-\infty}^\infty dt \sum_{\vec{x}},
 \end{eqnarray}
where $V=N_s a^3$ and the superscript $L$ denotes  they are lattice version of the continuum entities.  We are still in Minkowski spacetime. We keep the time continuous but dimensionless for convenience in the following discussion. The renormalization factor $Z_V$ for the lattice current $j^{L}_\mu=(\rho^L,-\vec{j}^L)$ can be taken to be unity if conserved currents are used on the lattice. Eq.\eqref{eq:4pt} becomes,
\begin{equation}
T_{\mu\nu}=i\ N_s a \int dt \sum_{\vec{x}} e^{ik_2\cdot x} \langl \pi(p_2) |T j^L_\mu(x) j^L_{\nu}(0) | \pi(p_1)\rangl.
\label{eq:4ptlat}
\end{equation}
On the lattice, there is a contribution to this function when $p_1=p_2$, called a vacuum expectation value (or VEV), that must be subtracted out. The reason is we are interested in differences relative to the vacuum, not the vacuum itself.
Formally, this is enforced by requiring normal ordering instead of time ordering in Eq.\eqref{eq:4ptlat},
\begin{equation}
:j^L_\mu(x) j^L_\nu(0) : = T j^L_\mu(x) j^L_\nu(0) - \langl 0|T j^L_\mu(x) j^L_\nu(0) | 0\rangl.
\label{eq:time}
\end{equation}
Including this subtracted contribution, the entire $\mu=0,\nu=0$ correlator may be characterized as
\begin{equation}
T_{00}  \equiv T^{elas}_{00} + T^{inel}_{00}.
\label{eq:T00_pion_lat}
\end{equation}
After insertion of a complete set of intermediate lattice states between the currents, the elastic part $T_{00}^{elas}$ is separated from the inelastic part.
It turns out the Born term $T_{00}^{Born}$ in the continuum cancels exactly the elastic term on the lattice. So the matching produces~\footnote{In this work we use $\vec{k}$ to denote continuum momentum and $\vec{q}$ lattice momentum with the same physical unit. When we match the two forms we set $\vec{k}=\vec{q}$ and express the result in terms of $\vec{q}$.}

\begin{equation}
T_{00}^{inel}(\vec{q}) = {\alpha_E \over \alpha}  \vec{q}^{\,2},
\end{equation}
or a formula for charged pion electric polarizability on the lattice,
\begin{equation}
\alpha^\pi_E=  {\alpha  \over  \vec{q}_1^2 } \left[ T_{00}(\vec{q}_1) - T_{00}^{elas}(\vec{q}_1) \right],
\label{eq:alpha_pion}
\end{equation}
where $\vec{q}_1$ emphasizes that the formula is valid for the smallest non-zero spatial momentum on the lattice.

Charged pion magnetic polarizability proceeds in a similar fashion, except we consider the spatial component $T_{11}$ ($T_{22}$ gives the same result). 
Under the same kinematics given  in Eq.\eqref{eq:Breit1}, 
this component from the general form in Eq.\eqref{eq:Tmn} reads
\begin{equation}
T_{11}= -{1\over m_\pi} + \vec{k}^{\,2} \left( {\langl r^2\rangl \over 3} + {\beta_M \over \alpha} \right).
\label{eq:4pt11}
\end{equation}
On the other hand Eqs.\eqref{eq:4ptlat} and \eqref{eq:time} for $\mu=1,\nu=1$ give the lattice form. Again, a complete set of lattice states are inserted between the currents. Unlike $T_{00}$, the elastic piece in the sum vanishes under the special kinematics, 
\begin{equation}
\langl\pi(\vec{0})| j^L_1(0) | \pi(\vec{q},s)\rangl=0,
\end{equation}
since the matrix element is proportional to $(\vec{0}+\vec{q})_1$ in 1-direction but momentum $\vec{q}$ is in 3-direction.

For the inelastic contributions, the types of intermediate state contributing are vector or axial vector mesons~\cite{Wilcox:1996vx}. There is no need to analyze the matrix elements explicitly as done in Ref.~\cite{Wilcox:1996vx} for the electric case. We only need to know that the inelastic part can be characterized up to order $\vec{q}^{\,2}$ by the form,
\begin{equation}
T_{11}(\vec{q}) \equiv T_{11}(\vec{0}) + \vec{q}^{\,2}  K_{11},
\label{eq:inelastic_lat}
\end{equation}
with $T_{11}(\vec{0})$ and $K_{11}$ to be related to physical parameters and determined on the lattice.
Note that we deliberately use the full amplitude label $T_{11}$ instead of $T^{inel}_{11}$ since the elastic part is zero.

Matching the full amplitude on the lattice in Eq.\eqref{eq:4pt11} with the continuum version in Eq.\eqref{eq:inelastic_lat}, 
we obtain two relations,
\begin{equation}
 -{1\over m_\pi}  = T_{11}(\vec{0}), \,\, 
  {\langl r^2\rangl \over 3m_\pi} + {\beta_M \over \alpha}  = K_{11}.  \label{eq:K11}
  \end{equation}
The first relation is a sum rule at zero momentum.
The second leads to a formula for charged pion magnetic polarizability, 
\begin{equation}
\beta^\pi_M= \alpha \left[- {\langl r^2\rangl \over 3m_\pi} + { T_{11}(\vec{q}_1) -  T_{11}(\vec{0})\over \vec{q}_1^2} \right],
\label{eq:beta_pion}
\end{equation}
where  $\vec{q}_1$ is the lowest momentum on the lattice.
Compared to charged pion electric polarizability $\alpha^\pi_E$ in Eq.\eqref{eq:alpha_pion}, we see that instead of subtracting the elastic contribution, we subtract the zero-momentum inelastic contribution in the magnetic polarizability. 
In other words, there is no zero-momentum contribution in $\alpha^\pi_E$, and no elastic contribution in $\beta^\pi_M$.

\section{Proton}
\label{sec:proton} 

We start with a unpolarized proton Compton tensor parameterized to second order in photon momentum,
\begin{eqnarray}
&\sqrt{2E_12E_2} \,T_{\mu\nu} = T_{\mu\nu}^{Born}
+ B(k_1\cdot k_2 g_{\mu\nu} -  k_{2\mu}k_{1\nu}) 
+ C(k_1\cdot k_2 Q_\mu Q_\nu + Q\cdot k_1 Q\cdot k_2  g_{\mu\nu} \label{eq:Tmn_p}\\
&\quad\quad - Q\cdot k_2 Q_\mu k_{1\nu} - Q\cdot k_1 Q_\nu k_{2\mu}),\nonumber
\end{eqnarray} 
where $Q=p_1+p_2$.
For Born term we take from Ref.~\cite{Gasser:2015dwa},
\begin{eqnarray}
T_{\mu\nu}^{Born} = {B_{\mu\nu}(p_2,k_2,s_2 | p_1,k_1,s_1) \over m_p^2 - s} 
+ {B_{\nu\mu}(p_2,-k_1,s_2 | p_1,-k_2,s_1) \over m_p^2 - u},
\label{eq:TBorn}
\end{eqnarray}
where the function is (note a factor of $1/2$ difference between our definition and Ref.~\cite{Gasser:2015dwa}),
\begin{eqnarray}
 B_{\mu\nu}(p_2,k_2,s_2 | p_1,k_1,s_1) = 
\bar{u}(p_2,s_2) \Gamma_{\mu}(-k_2) (\fsl{P}+m_p) \Gamma_{\nu}(k_1)  u(p_1,s_1).
\end{eqnarray}
Here $P=p_2+k_2=p_1+k_1$ is the standard 4-momentum conservation for Compton scattering. 
There is no $A$ term here because the proton Born terms obey current conservation, unlike the pion case in Eq.\eqref{eq:Tmn}.
The $B$ and $C$ are still related to polarizabilities as in Eq.\eqref{eq:BC}.

The Born amplitude has virtual (or off-shell) intermediate hadronic states in the s and u channels, whereas on the lattice we have real (or on-shell) intermediate states. This will produce a difference with the elastic contribution to be discussed later.
The vertex function is defined by
\begin{equation}
 \Gamma_{\mu}(k)=\gamma_\mu F_1 + {i F_2\over 2m_p} \sigma_{\mu\lambda} k^\lambda,
\end{equation}
where $F_{1,2}$ are the Dirac and Pauli form factors. Specializing to our kinematics in Eq.\eqref{eq:Breit1}, we have
\begin{equation}
s=(p_1+k_1)^2=m_p^2-\vec{k}^{\,2},
u=(p_1-k_2)^2=m_p^2-\vec{k}^{\,2}, 
\end{equation}
Including the contact interaction term, the generic amplitude has the form to order $\vec{k}^{\,2}$,
\begin{equation}
T_{00}(\vec{k}) = T_{00}^{Born}(\vec{k}) +  \vec{k}^{\,2} {\alpha^p_E \over \alpha}.
\end{equation}

On the other hand, we consider the unpolarized four-point function of the proton in lattice regularization,
\begin{align}
&T_{\mu\nu}=i\ N_s a {1\over 2} \sum_{s_1,s_2} 
\int_{-\infty}^\infty dt \sum_{\vec{x}} e^{ik_2\cdot x} 
\langl p_2,s_2|\left[ T j^L_\mu(x) j^L_\nu(0) 
- \langl 0 |T j^L_\mu(x) j^L_\nu(0) |0\rangl \right] |p_1,s_1\rangl, 
\label{eq:Tfull}
\end{align}
where the VEV subtraction is included. For electric polarizability, we are interested in the $\mu=\nu=0$ component of Eq.\eqref{eq:Tfull},
Matching the lattice and continuum forms and subtracting off the elastic contribution, we have
\begin{equation}
T_{00}(\vec{q}) - T_{00}^{elas}(\vec{q}) =  T_{00}^{Born}(\vec{q}) - T_{00}^{elas}(\vec{q}) + \vec{q}^{\,2} {\alpha^p_E \over \alpha}. 
\end{equation}
Many terms cancel between $T_{00}^{Born}$ and $T_{00}^{elas}$, leaving the difference,
\begin{equation}
T_{00}(\vec{q}) - T_{00}^{elas}(\vec{q}) = -{ (1+\kappa)^2 \over 4 m_p^3} \vec{q}^{\,2} 
+ {\alpha^p_E \over \alpha}  \vec{q}^{\,2},
\end{equation}
from which we arrive at a final formula for proton electric polarizability,
\begin{equation}
\alpha^p_E=  \alpha \left[ {(1+\kappa)^2 \over 4 m_p^3} 
+ {T_{00}(\vec{q}_1) - T_{00}^{elas}(\vec{q}_1) \over  \vec{q}_1^{\,2} } \right].
\label{eq:alpha_p}
\end{equation}

For the proton magnetic polarizability,
we start with the $\mu=\nu=1$ component of Eq.\eqref{eq:Tmn_p} (the 22 component gives the same result). Including the contact interaction term, the full amplitude can be characterized as
\begin{equation}
T_{11}(\vec{k}) = T_{11}^{Born}(\vec{k}) +  \vec{k}^{\,2} {\beta^p_E \over \alpha}.
\label{eq:T11_cont}
\end{equation}
Unlike the charged pion, there is an elastic contribution for the proton magnetic case. The inelastic 11 component in Eq.\eqref{eq:Tfull} can be formally characterized as a constant plus a linear term in $\vec{q}^{\,2}$, 
\begin{equation}
T_{11}^{inel}(\vec{q}) \equiv T_{11}^{inel}(\vec{0}) +  \vec{q}^{\,2} K_{11},
\label{eq:inelastic_lat_p}
\end{equation}
with $T_{11}^{inel}(\vec{0})$ and $K_{11}$ to be matched with physical parameters.
The difference between the Born term in the continuum and the elastic term on the lattice is ($\vec{k}\rightarrow \vec{q}$ in Born)
\begin{equation}
T_{11}^{Born}-T_{11}^{elas} = 
-{1 \over m_p} + \vec{q}^{\,2}  
\left(  {1 \over 2m_p^3} +  {\langl r_E^2\rangl \over 3m_p} \right),
\label{eq:T11diff}
\end{equation}
where the $\kappa$ terms in the zero-momentum part cancel, as well as the magnetic charge radius terms in the $\vec{q}^{\,2}$ part. By matching the full $T_{11}$ in the continuum and on the lattice, we have,
\begin{equation}
T_{11}^{elas}(\vec{q}) + T_{11}^{inel}(\vec{0}) +  \vec{q}^{\,2} K_{11} =T_{11}^{Born}(\vec{q}) +  \vec{q}^{\,2} {\beta^p_M\over \alpha}.
\end{equation}
Using Eq.\eqref{eq:T11diff}, we obtain two relations,
\begin{equation}
 T_{11}^{inel}(0) = -{1\over m_p}, 
\, K_{11} = {1 \over 2m_p^3} + {\langl r_E^2\rangl \over 3m_p}  + {\beta^p_M \over \alpha }.
 \label{eq:K11_proton}
\end{equation}
We see the same sum rule in the first relation as for the charged pion. The second relation produces an expression for proton magnetic polarizability on the lattice, 
\begin{equation}
\beta^p_M  = \alpha \left[- {1 \over 2m_p^3} -  {\langl r_E^2\rangl \over 3m_p}
+ { T^{inel}_{11}(\vec{q}_1) - T^{inel}_{11}(\vec{0}) \over \vec{q}_1^{\,2}} \right],
\end{equation}
where we have used Eq.\eqref{eq:inelastic_lat_p} for $K_{11}$. It turns out there is no elastic part to the zero momentum amplitude $T_{11}(\vec{0})$. 
Using the full amplitude $T_{11}$, we write the final lattice formula for proton magnetic polarizability as, 
\begin{eqnarray}
\beta^p_M  = \alpha \left[- {1 \over 2m_p^3} -  {\langl r_E^2\rangl \over 3m_p} \right.
 \left. + { T_{11}(\vec{q}_1) - T^{elas}_{11}(\vec{q}_1) - T_{11}(\vec{0}) \over \vec{q}_1^{\,2}} \right].
 \label{eq:beta_p}
\end{eqnarray}
Compared to charged pion magnetic polarizability $\beta^\pi_M$  in Eq.\eqref{eq:beta_pion}, proton $\beta^p_M$ has two extra terms: 
a mass contribution and an elastic contribution.

\section{Lattice measurement}
\label{sec:lat} 
Having obtained polarizability formulas in Eq.\eqref{eq:alpha_pion} and Eq.\eqref{eq:beta_pion} for charged pion,
and Eq.\eqref{eq:alpha_p} and Eq.\eqref{eq:beta_p} for proton,
we  now discuss how to measure them in lattice QCD.
First, we need to match the kinematics used in deriving the expressions, {\it i.e.}, with hadrons at rest and photons having spacelike momentum in the z-direction~\footnote{For general discussion, we use 
$h$ to represent either charged pion or proton.},
\begin{eqnarray}
&p_1 =(m_h,\vec{0}),\nonumber \\
&q_1 =(0,q\hat{z}),\;
q_2 =(0, -q\hat{z}), \; q \ll m_h,\\
&p_2 =q_2+q_1+p_1=(m_h,\vec{0}),\nonumber
\label{eq:Breit2}
\end{eqnarray}
It is the same kinematics as in Eq.~\eqref{eq:Breit1} but expressed differently to match what is being done on the lattice. We construct the four-point current-current correlation function,
\beq
P_{\mu\nu}(\vec{x}_2,\vec{x}_1,t_3,t_2,t_1,t_0) \equiv  
 {\langl 0 | \psi^\dagger (x_3) :j^L_\mu(x_2) j^L_\nu(x_1):  \psi (x_0) |0\rangl
\over 
 \langl 0 | \psi^\dagger (t_3) \psi (t_0) |0\rangl 
},
 \label{eq:P1}
\eeq
where the two-point function is for normalization, $\psi$ is the interpolating field of the hadron, and normal ordering is used to include the VEV contribution. In the case of proton, sum over final spin and average over initial spin are assumed for unpolarized measurement.
Note there is no explicit reference to $\vec{x}_3$ and $\vec{x}_0$ since they are absorbed in building zero-momentum hadrons at fixed times $t_3$ and $t_0$ from wall-sourced quark propagators. We have $t_3>t_{1,2}>t_0$ and $t_{1,2}$ indicates the two possibilities of time ordering.
When the times are well separated (defined by the time limits $t_3\gg t_{1,2} \gg t_0$) the correlator is dominated by
the ground state, 
\begin{equation}
P_{\mu\nu}(\vec{x}_2,\vec{x}_1,t_3,t_2,t_1,t_0)   \to \langl h(\vec{0}) | T j^L_\mu(r) j^L_\nu(0) |  h(\vec{0})\rangl 
 -  \langl 0 | T j^L_\mu(r) j^L_\nu(0) | 0\rangl,
 \end{equation}
where translation invariance has been used to shift the bilinear to $r=x_2-x_1$ and $0$. 
To implement the special kinematics we consider the Fourier transform (suppressing $t_3$ and $t_0$ for clarity)
\begin{equation}
Q_{\mu\nu}(\vec{q},t_2,t_1)  \equiv N_s\sum_{\vec{r}} e^{-i\vec{q}\cdot \vec{r}} 
P_{\mu\nu}(\vec{x}_2,\vec{x}_1,t_3,t_2,t_1,t_0) ,
\label{eq:Qmn}
\end{equation}
where $\vec{q}$ is lattice momentum. Charged pion electric polarizability in Eq.\eqref{eq:alpha_pion} is measured on the lattice by 
\begin{equation}
\alpha^{\pi}_E={2\alpha a \over \vec{q}_1^{\,2}} \int_{0}^\infty d t \left[Q_{00}(\vec{q}_1,t) -Q^{elas}_{00}(\vec{q}_1,t) \right].
\label{eq:alpha_pion_lat}
\end{equation}
Charged pion magnetic polarizability in Eq.\eqref{eq:beta_pion} is measured on the lattice by
\begin{equation}
\beta^{\pi}_M =\alpha \left\{- { \langl r_E^2\rangl \over 3m_\pi}
 +{2a\over  \vec{q}_1^{\,2}} \int_{0}^\infty d t \left[Q_{11}(\vec{q}_1,t) -Q_{11}(\vec{0},t)\right] \right\}, 
\label{eq:beta_pion_lat}
\end{equation}
where $Q_{11}(\vec{q}_1,t)$ is the 11 component of Eq.\eqref{eq:Qmn}. 
The four-point function $Q_{00}(\vec{q}_1,t) $ already contains information on $\langl r_E^2\rangl$ in its elastic limit~\cite{Wilcox_1992,Andersen:1996qb}. The proton electric polarizability in Eq.\eqref{eq:alpha_p} can be measured on the lattice by
\begin{align}
\alpha^{p}_E =\alpha  \left\{{(1+\kappa)^2 \over 4 m_p^3} \right. \label{eq:alpha_p_lat} 
 \left. +{2a\over \vec{q}_1^{\,2}} \int_{0}^\infty d t \left[Q_{00}(\vec{q}_1,t) -Q^{elas}_{00}(\vec{q}_1,t)\right] \right\},
\end{align}
and the magnetic polarizability in Eq.\eqref{eq:beta_p} by
\begin{align}
\beta^{p}_M =\alpha  \left\{-{1 \over 2m_p^3} -  {\langl r_E^2\rangl \over 3 m_p}  \right. \label{eq:beta_p_lat} 
 \left. +{2a\over  \vec{q}_1^{\,2}} \int_{0}^\infty d t \left[Q_{11}(\vec{q}_1,t) -Q^{elas}_{11}(\vec{q}_1,t) -Q_{11}(\vec{0},t)\right] \right\}. 
\end{align}
$Q_{11}^{elas}(\vec{q}_1,t)$ and $Q_{00}^{elas}(\vec{q}_1,t)$ contain the information to extract $\langl r_E^2\rangl$ and $(1+\kappa)^2 / 4 m_p^2$. The close coupling between the electric and magnetic suggests that it is most efficient to measure the two polarizabilities together, with associated mass, charge radius, and magnetic moment in the same simulation. This should be done on a configuration by configuration basis to maintain correlations.

\section{\label{sec:con} Conclusions and acknowledgements}

In this work we lay out a program for the use of four-point correlation functions 
by revitalizing an earlier study on electric polarizability of charged pions.
The approach bears a close resemblance to the physical Compton scattering process 
with a transparent physical picture and conceptual clarity.

Four-point function techniques are also useful for hadron structure function calculations leading to parton distribution functions. The same Compton meson and baryon quark-line diagrams are evaluated, except now at high momentum transfer. The key to this evaluation is the implementation of the inverse Laplace transform~\cite{Wilcox:1992xe}, such as the Bayesian reconstruction method employed in Ref.~\cite{Liang:2019frk}. Using this technique, useful comparisons on proposed continuum forms can be examined.

WW would like to acknowledge a Baylor University Arts and Sciences Research Leave. We also thank Xuan-He Wang and Yang Fu for catching an error in the intermediate steps for proton electric polarizability which did not affect the final results. This work was supported in part by DOE Grant~No.~DE-FG02-95ER40907.

\bibliographystyle{JHEP}
\bibliography{x4ptfun}

%

\end{document}